\documentclass[12pt]{article}
\usepackage{graphicx,amssymb,amsfonts,latexsym,amsmath,amsthm,times}
\usepackage{epsfig}
\usepackage{fancyhdr}
\usepackage{color}
\usepackage[english]{babel}

\numberwithin{equation}{section}
\begin{document}

{\footnotesize {\flushleft \mbox{\bf \textit{Math. Model. Nat.
Phenom.}}} \newline
\mbox{\textit{{\bf  2013 }}} }

{\footnotesize \thispagestyle{plain} }

{\footnotesize
\vspace*{2cm} }{\normalsize
\centerline{\Large \bf  Non-homogeneous random walks, subdiffusive
migration of cells } \centerline{\Large \bf  and anomalous chemotaxis } }

{\normalsize \vspace*{1cm}
}

{\normalsize
\centerline{\bf S. Fedotov$^a$\footnote{Corresponding
author. E-mail: sergei.fedotov@manchester.ac.uk}, A. O. Ivanov $^b$
 and A. Y. Zubarev $^b$ } }

{\normalsize \vspace*{0.5cm} }

{\normalsize
\centerline{$^a$ School of Mathematics, The University of Manchester,
Manchester, M13 9PL, UK} }

{\normalsize
\centerline{$^b$ Department of Mathematical Physics, Ural
Federal University, Ekaterinburg, 620083, Russia} }

{\normalsize
}

{\normalsize \vspace*{1cm}}

Abstract. This paper is concerned with a non-homogeneous in space and
non-local in time random walk model for anomalous subdiffusive transport of
cells. Starting with a Markov model involving a structured probability
density function, we derive the non-local in time master equation and
fractional equation for the probability of cell position. We derive the
fractional Fokker-Planck equation for the density of cells and apply this
equation to the anomalous chemotaxis problem. We show the structural
instability of fractional subdiffusive equation with respect to the partial
variations of anomalous exponent. We find the criteria under which the
anomalous aggregation of cells takes place in the semi-infinite domain.


\noindent AMS subject classification:


\vspace*{1cm}

\setcounter{equation}{0}

\section{Introduction}

Random cell movement plays a very important role in embryonic morphogenesis,
wound healing, cancer cells proliferation, and many other physiological and
pathological processes \cite{R}. The microscopic theory of the migration of
cells and bacteria towards a favorable environment (chemotaxis) is based on
random walk models \cite{EO,Hillen,Alt,OS1}. The \textquotedblleft
velocity-jump\textquotedblright\ models concern with self-propelled motion
involving the runs and tumbles, while \textquotedblleft
space-jump\textquotedblright models deal with the cells making jumps in
space. Much of the literature on the theoretical studies of cells motility
has been concerned with Markov random walk models (see, for example, \cite%
{Baker1,EO}). However, the analysis of random movement of wild-type and
mutated epithelial cells shows the anomalous dynamics of cell migration \cite%
{D} (see also \cite{M}). Over the past few years there have been several
attempts to model non-Markovian anomalous cell transport involving
subdiffusion and superdiffusion \cite{D,Fed0,Fed00,Fed1,Fed2,HL}. In this
paper we shall deal with a non-Markovian \textquotedblleft
space-jump\textquotedblright model that describes the non-homogeneous in
space subdiffusive transport of cells.

\subsection{Markov random walk model.}

First let us consider a Markov model for random cell movement along
one-dimensional lattice such that all steps are of equal length $1$. We
define the probability%
\begin{equation}
p(k,t)=\Pr \left\{ X(t)=k\right\}
\end{equation}%
that the position of cell $X(t)$\ is at point $k\in \mathbb{Z}$\ at time $t.$%
We introduce at each point $k$\ the rate of jump to the left $\mu (k)$\ and
the rate of jump to the right $\lambda (k)$. This random walk is called a
generalized birth-death process \cite{Cox}. The master equation for $p(k,t) $%
\ can be written as
\begin{equation}
\frac{\partial p(k,t)}{\partial t}=\lambda (k-1)p(k-1,t)+\mu
(k+1)p(k+1,t)-\left( \lambda (k)+\mu (k)\right) p(k,t).  \label{Master11}
\end{equation}%
This model corresponds to the case when intervals between jumps at point $k$
are exponentially distributed with parameter $\lambda (k)+\mu (k).$ When the
cell makes a jump from the position $k,$\ it jumps to the right with
probability $\lambda (k)/(\lambda (k)+\mu (k))$ and to the left with the
probability $\mu (k)/(\lambda (k)+\mu (k))$ \cite{Cox}.

The dependence of $\mu (k)$ and $\lambda (k)$ on space can be introduced in
different ways depending on how cells sense the surrounding environment. For
the local chemotaxis models, the rates $\lambda (S(k))$\ and $\mu (S(k))$
are the functions of the local concentration of the chemotactic substance $%
S(k).$ There exist several non-local and barrier models that are different
in terms of the dependence of rate functions on the chemotactic substance
\cite{Baker1,OS1}. For example, the rates $\mu (k)$ and $\lambda (k)$ can
depend on the concentration of the chemotactic substance at neighbouring
positions $k-1$ and $k+1$ as in (\ref{nonlocal}). In the continuous limit,
the master equation (\ref{Master11}) can be reduced to the classical
advection-diffusion equation in which the cell flux involves the standard
diffusion term and the advection term due to chemotaxis.

If we consider only positive values of $k,$ we need to implement boundary
conditions at the point $k=1$. Here we assume that if cell hits the wall on
the boundary, it is reflected with the probability $1-\chi $ and absorbed by
wall with the probability $\chi .$ Then one can write $p(1,t+\Delta
t)=(1-\lambda (1)\Delta t-\mu (1)\Delta t)p(1,t)+$ $\mu (1)(1-\chi
)p(1,t)\Delta t+\mu (2)p(2,t)\Delta t+o(\Delta t).$ In the limit $\Delta
t\rightarrow 0$ we obtain
\begin{equation}
\frac{\partial p(1,t)}{\partial t}=-\chi \mu (1)p(1,t)+\mu (2)p(2,t)-\lambda
(1)p(1,t),  \label{bound}
\end{equation}%
where $0\leq \chi \leq 1.$

Non-uniform stationary solution of master equation (\ref{Master11}) can be
interpreted as cell aggregation phenomenon \cite{OS1}. In particular, if
there is no absorption on the boundary ($\chi =0$), the stationary solution $%
p_{st}(k)$ can be easily found from (\ref{Master11}) and (\ref{bound}). We
obtain
\begin{equation}
p_{st}(k)=p_{st}(1)\prod\limits_{i=1}^{k-1}\frac{\lambda (i)}{\mu (i+1)}%
,\qquad k>1,  \label{ma}
\end{equation}%
where%
\begin{equation*}
p_{st}(1)=\left( 1+\sum\limits_{k=2}^{\infty }\prod\limits_{i=1}^{k-1}\frac{%
\lambda (i)}{\mu (i+1)}\right) ^{-1}
\end{equation*}%
provided the series is convergent.

\subsection{Anomalous random walks}

It is tempting to generalize the master equation (\ref{Master11}) for the
anomalous case by replacing the time derivative with the Caputo derivative
\cite{Klages,Mee,Met2}
\begin{equation}
\frac{\partial ^{\nu }p\left( k,t\right) }{\partial t^{\nu }}=\frac{1}{%
\Gamma (1-\nu )}\int_{0}^{t}\frac{\partial p\left( k,u\right) }{\partial u}%
\frac{du}{(t-u)^{1-\nu }}  \label{Caputo}
\end{equation}%
as it is done in \cite{FBD} for a fractional linear birth--death process.
Here $\nu $ is the anomalous exponent: $0<\nu <1.$ Although this
generalization is very attractive from a mathematical point of view, it is
not appropriate for a non-homogeneous medium for which the exponent $\nu $
depends on $k$. The non-homogeneous fractional equation for $p(k,t)$ can be
written as
\begin{equation}
\frac{\partial p(k,t)}{\partial t}=a(k-1)\mathcal{D}_{t}^{1-\nu
(k-1)}p(k-1,t)+b(k+1)\mathcal{D}_{t}^{1-\nu (k+1)}p(k+1,t)-(a(k)+b(k))%
\mathcal{D}_{t}^{1-\nu (k)}p(k,t),  \label{dis1}
\end{equation}%
where \bigskip $\mathcal{D}_{t}^{1-\nu (k)}$ is the Riemann-Liouville
fractional derivative with varying order%
\begin{equation}
\mathcal{D}_{t}^{1-\nu (k)}p\left( k,t\right) =\frac{1}{\Gamma (\nu (k))}%
\frac{\partial }{\partial t}\int_{0}^{t}\frac{p\left( k,u\right) du}{%
(t-u)^{1-\nu (k)}}.  \label{RL}
\end{equation}%
Here $\nu (k)$ is the anomalous exponent corresponding to the site $k$ and
the anomalous rate coefficients $a(k)$ and $b(k)$ have to be determined, see
(\ref{ab}). The crucial point here is that the anomalous exponent $\nu (k)$
depends on the site $k$. The fractional equation (\ref{dis1}) cannot be
rewritten in terms of Caputo derivative (\ref{Caputo}). It turns out that
even small non-homogeneous variations of the exponent $\nu $ lead to a
drastic change of $p(k,t)$ in the limit $t\rightarrow \infty $ \ \cite{Fed3}%
. It means that the subdiffusive fractional equations with constant
anomalous exponent $\nu $ are\textit{\ not structurally stable}. If, for
example, the point $k=M$ has the property that $\nu (M)<\nu (k)$ for all $%
k\neq M$ and $\chi =0,$ one can find that
\begin{equation}
p(k,t)\rightarrow 0,\qquad p(M,t)\rightarrow 1,\qquad 1\leq k\leq N
\end{equation}%
as $t\rightarrow \infty .$ This result has been interpreted as anomalous
aggregation of cells at the point $k=M$ \cite{Fed1}. In this paper we shall
find the conditions for anomalous aggregation for the semi-infinite interval
$1\leq k<\infty .$ It should also be noted that non-homogeneous variations
of the exponent $\nu $ destroy the Gibbs-Boltzmann distribution as a long
time limit of the fractional Fokker-Planck equation \cite{Fed3}. Of course,
for the constant value of $\nu ,$ the formulation in terms of Caputo or
Riemann-Liouville operators are equivalent, as long as proper care is taken
of the initial values \cite{Klages,Mee,Met2}.

\subsection{Anomalous diffusion with reaction.}

Another extension of traditional Markov random walks models is non-Markovian
theory of anomalous transport with reaction dynamics \cite%
{Fed000,MFH,Nep,Sokolov,Sh,Vo}. In particular, this theory has been used for
the analysis of the proliferation and migration dichotomy of cancer cells
\cite{Fed0,Fed00,Fed2}. In this paper we consider the inhibition of cell
growth by anticancer therapeutic agents. To model this inhibition we
introduce the random death process with non-uniform death rate parameter. We
assume that during time interval $\left( t,t+\Delta t\right) $ at point $k$\
each cell has a chance $\theta (k)\Delta t+o(\Delta t)$ of dying, where $%
\theta (k)$ is the death rate \cite{Iomin}. It is easy to take into account
this process for Markov models. We just add the term $-\theta (k)p(k,t)$ to
the right hand side of the master equation (\ref{Master11}). On the
contrary, the anomalous master equation involves a non-trivial combination
of transport and death kinetic terms because of memory effects \cite%
{HLW,MFH,Abad}. In this paper we shall derive the following fractional
equation
\begin{eqnarray}
\frac{\partial p(k,t)}{\partial t} &=&a(k-1)e^{-\theta (k-1)t}\mathcal{D}%
_{t}^{1-\nu (k-1)}\left[ p(k-1,t)e^{\theta (k-1)t}\right]  \notag \\
&&+b(k+1)e^{-\theta (k+1)t}\mathcal{D}_{t}^{1-\nu (k+1)}\left[
p(k+1,t)e^{\theta (k+1)t}\right]  \notag \\
&&-\left( a(k)+b(k)\right) e^{-\theta (k)t}\mathcal{D}_{t}^{1-\nu (k)}\left[
p(k,t)e^{\theta (k)t}\right] -\theta (k)p(k,t).  \label{master30}
\end{eqnarray}

\subsection{Mean field master equation for the density of cells.}

Instead of the probability $p(k,t)$ for an individual cell one can consider
the mean density of cells $\rho (x,t)$ as a function of space $x$ and time $%
t $. The master equation (\ref{Master11}) can be rewritten as the equation
for the density $\rho (x,t)$ by changing the variables as $k\rightarrow x$
and $k\pm 1\rightarrow x\pm l$:
\begin{equation}
\frac{\partial \rho (x,t)}{\partial t}=\lambda (x-l)\rho (x-l,t)+\mu
(x+l)\rho (x+l,t)-(\lambda (x)+\mu (x))\rho (x,t)-\theta (x)\rho (x,t),
\label{Master10}
\end{equation}%
where $l$ is the jump size, $\theta (x)$ is the death rate. The advantage of
this equation is that one can easily take into account various non-linear
effects by assuming the dependence of the rate functions $\lambda (x),\mu
(x) $ and $\theta (x)$ on the average density $\rho (x,t).$

In the anomalous subdiffusive case, the master equation for mean field $\rho
(x,t)$ can be obtained from (\ref{master30}). It can be written as a mass
balance equation
\begin{equation}
\frac{\partial \rho (x,t)}{\partial t}=-I(x,t)+I(x-l,t)-\theta (x)\rho (x,t),
\label{balance}
\end{equation}%
where $I(x,t)$ is the total flow of cells from the point $x$ to $x+l$
\begin{equation}
I(x,t)=a(x)e^{-\theta (x)t}\mathcal{D}_{t}^{1-\nu (x)}\left[ e^{\theta
(x)t}\rho (x,t)\right] -b(x+l)e^{-\theta (x+l)t}\mathcal{D}_{t}^{1-\nu (x+l)}%
\left[ e^{\theta (x+l)t}\rho (x+l,t)\right]  \label{flux}
\end{equation}%
and $I(x-l,t)$ is the total flow of cells from the point $x-l$ to $x$
\begin{equation}
I(x-l,t)=a(x-l)e^{-\theta (x-l)t}\mathcal{D}_{t}^{1-\nu (x-l)}\left[
e^{\theta (x-l)t}\rho (x-l,t)\right] -b(x)e^{-\theta (x)t}\mathcal{D}%
_{t}^{1-\nu (x)}\left[ e^{\theta (x)t}\rho (x,t)\right] .
\end{equation}%
Here $a(x)$ and $b(x)$ are the anomalous rate functions, see (\ref{ab}). One
can see that the flow of cells $I(x,t)$ depends on the death rate $\theta
(x) $. It means that in the anomalous case one cannot separate the transport
of cells from the death process \cite{HLW}. This phenomenon does not exist
in the Markovian case. For the Markov model (\ref{Master10}) the flux $I(x,t)
$ is independent from $\theta (x):$
\begin{equation}
I(x,t)=\lambda (x)\rho (x,t)-\mu (x+l)\rho (x+l,t).  \notag
\end{equation}%
When the density $\rho (x,t)$ is conserved ($\theta =0$), the master
equation (\ref{balance}) can be approximated by the fractional Fokker-Planck
equation \cite{Klages,Met1,Met2}%
\begin{equation}
\frac{\partial \rho (x,t)}{\partial t}=-\frac{\partial }{\partial x}\left[
l(a(x)-b(x))\mathcal{D}_{t}^{1-\nu (x)}\rho (x,t)\right] +\frac{\partial ^{2}%
}{\partial x^{2}}\left[ \frac{l^{2}}{2}(a(x)+b(x))\mathcal{D}_{t}^{1-\nu
(x)}\rho (x,t)\right] .  \label{FFP}
\end{equation}%
This is an example of the fractional equation with varying anomalous
exponent \cite{Ch}. Note that $a(x)-b(x)\sim l$ as $l\rightarrow 0$, see (%
\ref{dif2}).

The purpose of the next section is to set up a non-Markovian discrete-space
random walk model describing cell motility involving memory effects, the
death process and subdiffusive transport.

\section{Non-Markovian discrete-space random walk model}

\subsection{ Random cell motility}

There exist numerous mechanisms that facilitate random cell movement \cite{R}%
. In this paper we adopt the following random model of cell motility. When
the cell makes a jump to position $k$, the time the cell spends here before
it makes a jump to point $k-1$ or $k+1$ is random. It is called the
residence time or waiting (holding) time. We define the residence time at
position $k$ as
\begin{equation}
T_{k}=\min \left( T_{k}^{\mu },T_{k}^{\lambda }\right) ,  \label{min}
\end{equation}%
where $T_{k}^{\mu }$ and $T_{k}^{\lambda }$ are the independent random times
of jump to the left and right respectively. The idea here is that there
exist internal cellular signals involving two "hidden" independent\ random
alarm clocks. If one of the clocks goes off first, say $T_{k}^{\lambda
}<T_{k}^{\mu }$, the cell moves to the right to the point $k+1$. The other
clock "tells" the cell to move left to the point $k-1$ if it goes off first (%
$T_{k}^{\mu }<T_{k}^{\lambda })$. Note that migration of cells is a highly
complicated dynamic process which is regulated by both intercellular signals
and the surrounding environment. Since we do not know the exact mechanism of
cell motility we use a stochastic approach involving two random times $%
T_{k}^{\mu }$ and $T_{k}^{\lambda }$ for jumping to the left and right. Note
that if the random times $T_{k}^{\mu }$ and $T_{k}^{\lambda }$ are
exponentially distributed with the rates $\mu (k)$ and $\lambda (k)$
respectively, we have a classical Markov model with the master equation (\ref%
{Master11}). If the random variables $T_{k}^{\mu }$ and $T_{k}^{\lambda }$
are not exponentially distributed, the standard Markov approach does not
work. In this section we consider the non-Markovian case when $T_{k}^{\mu }$
and $T_{k}^{\lambda }$ are independent positive random variables with
general survival functions
\begin{equation}
\Psi _{\mu }(k,\tau )=\Pr \left\{ T_{k}^{\mu }>\tau \right\} ,\qquad \Psi
_{\lambda }(k,\tau )=\Pr \left\{ T_{k}^{\lambda }>\tau \right\} .
\label{sur55}
\end{equation}%
The Markov model (\ref{Master11}) corresponds to the following choice
\begin{equation}
\Psi _{\mu }(k,\tau )=e^{-\mu (k)\tau },\qquad \Psi _{\lambda }(k,\tau
)=e^{-\lambda (k)\tau }.
\end{equation}%
It is convenient to introduce the rate of escape (hazard function) $\gamma
(k,\tau )$ from the point $k$ as
\begin{equation}
\gamma (k,\tau )=\lim_{h\rightarrow 0}\frac{\Pr \left\{ \tau <T_{k}<\tau
+h\;|_{T_{k}>\tau }\right\} }{h}.  \label{tran}
\end{equation}%
If we denote the survival function at the point $k$ as
\begin{equation*}
\Psi (k,\tau )=\Pr \left\{ T_{k}>\tau \right\}
\end{equation*}%
and the residence time probability density function as%
\begin{equation*}
\psi (k,\tau )=-\frac{\partial \Psi (k,\tau )}{\partial \tau },
\end{equation*}%
then \cite{Cox}
\begin{equation}
\gamma (k,\tau )=\frac{\psi (k,\tau )}{\Psi (k,\tau )}.  \label{def}
\end{equation}%
Now we determine this rate function in terms of statistical characteristics
of random residence times $T_{k}^{\mu }$ and $T_{k}^{\lambda }.$ It follows
from the definition of the residence time $T_{k}$ at position $k$ (\ref{min}%
) that the survival function $\Psi (k,\tau )$ can be written as a product%
\begin{equation*}
\Psi (k,\tau )=\Psi _{\lambda }(k,\tau )\Psi _{\mu }(k,\tau ),
\end{equation*}%
where $\Psi _{\lambda }(k,\tau )$ and $\Psi _{\mu }(k,\tau )$ are defined by
(\ref{sur55}). Differentiation of this equation with respect to $\tau $ gives%
\begin{equation}
\psi (k,\tau )=\psi _{\lambda }(k,\tau )+\psi _{\mu }(k,\tau ),  \label{wtd}
\end{equation}%
where the transition densities  $\psi _{\lambda }(k,\tau
) $ and $\psi _{\mu }(k,\tau )$ are defined as
\begin{equation}
\psi _{\lambda }(k,\tau )=-\frac{\partial \Psi _{\lambda }(k,\tau )}{%
\partial \tau }\Psi _{\mu }(k,\tau ),\qquad \psi _{\mu }(k,\tau )=-\frac{%
\partial \Psi _{\mu }(k,\tau )}{\partial \tau }\Psi _{\lambda }(k,\tau ).
\end{equation}%
The formula (\ref{wtd}) is the particular case of the general expression for
the residence time PDF in terms of the transition densities (see
formula (5) in the classical paper by van Kampen \cite{Kampen}). These
transition densities have a clear probabilistic meaning. For example, $\psi
_{\mu }(k,\tau )\Delta \tau $ is the probability that the cell's jump to the
left occurs in the time interval $(\tau ,\tau +\Delta \tau )$ since the cell
arrived at point $k$. If we divide both sides of (\ref{wtd}) by the survival
function $\Psi (k,\tau )$ and use the formula (\ref{def}), we obtain
\begin{equation}
\gamma (k,\tau )=\lambda (k,\tau )+\mu (k,\tau ),  \label{rate}
\end{equation}%
where the rate of jump to the right $\lambda (k,\tau )$ and the rate of jump
to the left $\mu (k,\tau )$ are defined as
\begin{equation}
\lambda (k,\tau )=\frac{\psi _{\lambda }(k,\tau )}{\Psi (k,\tau )},\qquad
\mu (k,\tau )=\frac{\psi _{\mu }(k,\tau )}{\Psi (k,\tau )}.  \label{sur5}
\end{equation}%
Note that the transition rates $\lambda (k,\tau )$ and $\mu (k,\tau )$ can
be introduced from the very beginning as it is done in \cite{Kampen}. By
using (\ref{def}) and (\ref{rate}), we write the survival function $\Psi
(k,\tau )$ as
\begin{equation}
\Psi (k,\tau )=e^{-\int_{0}^{\tau }\lambda (k,\tau )d\tau -\int_{0}^{\tau
}\mu (k,\tau )d\tau }.  \label{s3}
\end{equation}%
The residence time probability density function $\psi (k,\tau )$ \bigskip
takes the form
\begin{equation*}
\psi (k,\tau )=(\lambda (k,\tau )+\mu (k,\tau ))e^{-\int_{0}^{\tau }\lambda
(k,\tau )d\tau -\int_{0}^{\tau }\mu (k,\tau )d\tau }.
\end{equation*}%
For the Markov case for which $\lambda (k)$ and $\mu (k)$ are independent of
the residence time variable $\tau ,$ we obtain from (\ref{s3}) the standard
exponential survival function
\begin{equation*}
\Psi (k,\tau )=e^{-\lambda (k)\tau -\mu (k)\tau }
\end{equation*}%
corresponding to the Markov master equation (\ref{Master11}).

\subsection{ Structured probability density function}

If the residence time probability density function $\psi $ is not
exponential, the random walk is non-Markovian. The standard method to deal
with non-Markovian stochastic processes is to add auxiliary variables to the
definition of the random walk to make the process Markovian \cite{Cox}. Here
we introduce the structured probability density function $\xi (k,t,\tau )$
involving residence time $\tau $ as auxiliary variable. The structural
density gives the probability that the cell position $X(t)$ at time $t$ is
at the point $k$ and its residence time $T_{k}$ at point $k$ is in the
interval $(\tau ,\tau +d\tau ).$ This is a standard way to deal with
non-Markovian random walks \cite{Cox,MFH}. Suppose that cells die at random
at rate $\theta (k)$ that depends on $k.$ The density $\xi (k,t,\tau )$
obeys the balance equation
\begin{equation}
\frac{\partial \xi }{\partial t}+\frac{\partial \xi }{\partial \tau }%
=-\lambda (k,\tau )\xi -\mu (k,\tau )\xi -\theta (k)\xi .  \label{basic}
\end{equation}%
We consider only the case when the residence time of random walker at $t=0$
is equal to zero, so the initial condition is
\begin{equation}
\xi (k,0,\tau )=p_{0}(k)\delta (\tau ),  \label{initial}
\end{equation}%
where $p_{0}(k)=$ $\Pr \left\{ X(0)=k\right\} $. The boundary condition in
terms of residence time variable ($\tau =0)$ can be written as \cite{Cox}
\begin{equation}
\xi (k,t,0)=\int_{0}^{t}\lambda (k-1,\tau )\xi (k-1,t,\tau )d\tau
+\int_{0}^{t}\mu (k+1,\tau )\xi (k+1,t,\tau )d\tau .  \label{arr}
\end{equation}%
In what follows we consider only positive values of $k$. In which case, we
have to specify the boundary condition for $k=1$. We write
\begin{equation}
\xi (1,t,0)=(1-\chi )\int_{0}^{t}\mu (1,\tau )\xi (1,t,\tau )d\tau
+\int_{0}^{t}\mu (2,\tau )\xi (2,t,\tau )d\tau .  \label{k=0}
\end{equation}%
This equation tells us that when cells escape from the point $k=1$ and move
to the left with the rate $\mu (1,\tau )$, they are adsorbed by the wall
with probability $\chi $, and reflected back to the position $k=1$ with the
probability $1-\chi .$ Note that this boundary condition can be written in
many different ways, for example, the cells can be reflected to state $k=2$.
One can also introduce a residence time PDF for a wall such that the
reflection is not instantaneous.

We solve (\ref{basic}) by the method of characteristics%
\begin{equation}
\xi (k,t,\tau )=\xi (k,t-\tau ,0)e^{-\int_{0}^{\tau }\lambda (k,\tau )d\tau
-\int_{0}^{\tau }\mu (k,\tau )d\tau -\theta (k)\tau },\quad \tau <t,\quad
k\geq 1.  \label{s1}
\end{equation}%
The structural density $\xi $ can be rewritten in terms of the survival
function $\Psi (k,\tau )$ (\ref{s3}) and the integral arrival rate
\begin{equation*}
j(k,t)=\xi (k,t,0)
\end{equation*}%
as%
\begin{equation}
\xi (k,t,\tau )=j\left( k,t-\tau \right) \Psi (k,\tau )e^{-\theta (k)\tau
},\quad \tau <t,\quad k\geq 1.  \label{je}
\end{equation}%
Our purpose now is to derive the master equation for the probability $%
p(k,t)=\Pr \left\{ X(t)=k\right\} :$%
\begin{equation}
p(k,t)=\int_{0}^{t}\xi (k,t,\tau )d\tau ,\quad k\geq 1.  \label{denG}
\end{equation}%
Let us introduce the integral escape rate to the right $i_{\lambda }(k,t)$
and the integral escape rate to the left $i_{\mu }(k,t)$ as%
\begin{equation}
i_{\lambda }(k,t)=\int_{0}^{t}\lambda (k,\tau )\xi (k,t,\tau )d\tau ,\qquad
i_{\mu }(k,t)=\int_{0}^{t}\mu (k,\tau )\xi (k,t,\tau )d\tau .  \label{i1}
\end{equation}%
Then the boundary conditions (\ref{arr}) and (\ref{k=0}) can be written in a
very simple form:
\begin{equation}
j(k,t)=i_{\lambda }(k-1,t)+i_{\mu }(k+1,t),\quad k>1  \label{j}
\end{equation}%
and
\begin{equation}
j(1,t)=(1-\chi )i_{\mu }(1,t)+i_{\mu }(2,t).
\end{equation}%
It follows from (\ref{sur5}), (\ref{initial}), (\ref{je}) and (\ref{i1})
that
\begin{equation}
i_{\lambda }(k,t)=\int_{0}^{t}\psi _{\lambda }(k,\tau )j(k,t-\tau
)e^{-\theta (k)\tau }d\tau +\psi _{\lambda }(k,t)p_{0}(k)e^{-\theta (k)t},
\label{i5}
\end{equation}%
\begin{equation}
i_{\mu }(k,t)=\int_{0}^{t}\psi _{\mu }(k,\tau )j(k,t-\tau )e^{-\theta
(k)\tau }d\tau +\psi _{\mu }(k,t)p_{0}(k)e^{-\theta (k)t}.  \label{i6}
\end{equation}%
Substitution of (\ref{initial}) and (\ref{je}) to (\ref{denG}), gives
\begin{equation}
p(k,t)=\int_{0}^{t}\Psi (k,\tau )j(k,t-\tau )e^{-\theta (k)\tau }d\tau +\Psi
(k,t)p_{0}(k)e^{-\theta (k)t}.  \label{p1}
\end{equation}%
It is convenient to introduce the integral escape rate $i(k,t)$ as the sum
of the escape rate to the right $i_{\lambda }(k,t)$ and the escape rate to
the left $i_{\mu }(k,t)$
\begin{equation}
i(k,t)=i_{\lambda }(k,t)+i_{\mu }(k,t).  \label{bau1}
\end{equation}%
The balance equation for $p(k,t)$ can be written as%
\begin{equation}
\frac{\partial p(k,t)}{\partial t}=-i(k,t)+j(k,t)-\theta (k)p(k,t),\quad k>1.
\label{balan}
\end{equation}%
To obtain a closed equation for $p(k,t)$ we need to express $i(k,t)$ and $%
j(k,t)$ in terms of $p(k,t).$ By applying the Laplace transform $\hat{\psi}%
(k,s)=\int_{0}^{\infty }\psi (k,\tau )e^{-s\tau }d\tau $ to (\ref{i5}), (\ref%
{i6}) and (\ref{p1}), we obtain
\begin{equation*}
\hat{\imath}_{\lambda }(k,s)=\hat{\psi}_{\lambda }(k,s+\theta (k))\left[
\hat{\jmath}(k,s)+p_{0}(k)\right] ,
\end{equation*}%
\begin{equation*}
\hat{\imath}_{\mu }(k,s)=\hat{\psi}_{\mu }(k,s+\theta (k))\left[ \hat{\jmath}%
(k,s)+p_{0}(k)\right]
\end{equation*}%
and%
\begin{equation*}
\hat{p}(k,s)=\hat{\Psi}(k,s+\theta (k))\left[ \hat{\jmath}(k,s)+p_{0}(k)%
\right] .
\end{equation*}%
In the Laplace space we have the following expressions for escape rates
\begin{equation}
\hat{\imath}_{\lambda }(k,s)=\frac{\hat{\psi}_{\lambda }(k,s+\theta (k))}{%
\hat{\Psi}(k,s+\theta (k))}\hat{p}(k,s),\qquad \hat{\imath}_{\mu }(k,s)=%
\frac{\hat{\psi}_{\mu }(k,s+\theta (k))}{\hat{\Psi}(k,s+\theta (k))}\hat{p}%
(k,s).
\end{equation}%
Using inverse Laplace transform and shift theorem we find
\begin{eqnarray}
i_{\lambda }(k,t) &=&\int_{0}^{t}K_{\lambda }(k,t-\tau )e^{-\theta
(k)(t-\tau )}p(k,\tau )d\tau ,  \notag \\
i_{\mu }(k,t) &=&\int_{0}^{t}K_{\mu }(k,t-\tau )e^{-\theta (k)(t-\tau
)}p(k,\tau )d\tau ,  \label{iii}
\end{eqnarray}%
where $K_{\lambda }(k,t)$ and $K_{\mu }(k,t)$ are the memory kernels defined
by Laplace transforms
\begin{equation}
\hat{K}_{\lambda }\left( k,s\right) =\frac{\hat{\psi}_{\lambda }(k,s)}{\hat{%
\Psi}\left( k,s\right) },\qquad \hat{K}_{\mu }\left( k,s\right) =\frac{\hat{%
\psi}_{\mu }(k,s)}{\hat{\Psi}\left( k,s\right) }.  \label{new5}
\end{equation}%
It follows from (\ref{j}), (\ref{bau1}), (\ref{balan}) and (\ref{iii}) that
the master equation for the probability $p(k,t)$ is%
\begin{eqnarray}
\frac{\partial p(k,t)}{\partial t} &=&\int_{0}^{t}K_{\lambda }(k-1,t-\tau
)p(k-1,\tau )e^{-\theta (k-1)(t-\tau )}d\tau  \notag \\
&&+\int_{0}^{t}K_{\mu }(k+1,t-\tau )p(k+1,\tau )e^{-\theta (k+1)(t-\tau
)}d\tau  \notag \\
&&-\int_{0}^{t}[K_{\lambda }(k,t-\tau )+K_{\mu }(k,t-\tau )]p(k,\tau
)e^{-\theta (k)(t-\tau )}d\tau -\theta (k)p(k,t)  \label{master99}
\end{eqnarray}%
for $k>1.$ The balance equation for $k=1$ is%
\begin{equation}
\frac{\partial p(1,t)}{\partial t}=-\chi i_{\mu }(1,t)-i_{\lambda
}(1,t)+i_{\mu }(2,t)-\theta (1)p(1,t)  \notag
\end{equation}%
or%
\begin{eqnarray}
\frac{\partial p(1,t)}{\partial t} &=&-\chi \int_{0}^{t}K_{\mu }(1,t-\tau
)p(1,\tau )e^{-\theta (1)(t-\tau )}d\tau -\int_{0}^{t}K_{\lambda }(1,t-\tau
)p(1,\tau )e^{-\theta (1)(t-\tau )}d\tau  \notag \\
&&+\int_{0}^{t}K_{\mu }(2,t-\tau )p(2,\tau )e^{-\theta (2)(t-\tau )}d\tau
-\theta (1)p(1,t),
\end{eqnarray}%
where $0\leq \chi \leq 1.$ The master equation for $p(k,t)$ can be rewritten
in terms of the probability flux $I(k,t)$ from the point $k$ to $k+1$
\begin{equation}
I(k,t)=\int_{0}^{t}K_{\lambda }(k,t-\tau )p(k,\tau )e^{-\theta (k)(t-\tau
)}d\tau -\int_{0}^{t}K_{\mu }(k+1,t-\tau )p(k+1,\tau )e^{-\theta
(k+1)(t-\tau )}d\tau  \label{I}
\end{equation}%
as
\begin{equation}
\frac{\partial p(k,t)}{\partial t}=-I(k,t)+I(k-1,t)-\theta (k)p(k,t).
\end{equation}%
In the next section we shall derive fractional master equation for $p(k,t).$

\section{\ Anomalous subdiffusion in heterogeneous media}

We now turn to the anomalous subdiffusive case. We assume that the longer
cell survives at point $k$, the smaller the transition probability from $k$
becomes. It means that the transition rates $\lambda (k,\tau )$ and $\mu
(k,\tau )$ are decreasing functions of residence time $\tau .$ We assume
that
\begin{equation}
\lambda (k,\tau )=\frac{\nu _{\lambda }(k)}{\tau _{0}(k)+\tau },\qquad \mu
(k,\tau )=\frac{\nu _{\mu }(k)}{\tau _{0}(k)+\tau },  \label{rates}
\end{equation}%
where $\tau _{0}(k)$ is a parameter with units of time. Both $\nu _{\lambda
}(k)$ and $\nu _{\mu }(k)$ play a very important role in what follows. From (%
\ref{s3}) and (\ref{rates}) we find that the survival function has a
power-law dependence
\begin{equation*}
\Psi (k,\tau )=\left( \frac{\tau _{0}(k)}{\tau _{0}(k)+\tau }\right) ^{\nu
(k)},
\end{equation*}%
where the exponent%
\begin{equation}
\nu (k)=\nu _{\lambda }(k)+\nu _{\mu }(k)  \label{exp}
\end{equation}%
depends on the state $k.$ Residence time probability density function $\psi
(k,\tau )=-\partial \Psi (k,\tau )/\partial \tau $ has the Pareto form
\begin{equation}
\psi (k,\tau )=\frac{\nu (k)\left( \tau _{0}(k)\right) ^{\nu (k)}}{(\tau
_{0}(k)+\tau )^{1+\nu (k)}},  \label{Pareto}
\end{equation}%
The anomalous subdiffusive case \cite{Klages,MFH} corresponds to%
\begin{equation*}
\nu (k)=\nu _{\lambda }(k)+\nu _{\mu }(k)<1.
\end{equation*}%
We can notice from (\ref{rates}) that the ratios $\lambda (k,\tau )$ and $%
\mu (k,\tau )$ to $\lambda (k,\tau )+\mu (k,\tau )$ are independent of the
residence time variable $\tau $ that is
\begin{equation*}
\frac{\lambda (k,\tau )}{\lambda (k,\tau )+\mu (k,\tau )}=\frac{\nu
_{\lambda }(k)}{\nu _{\lambda }(k)+\nu _{\mu }(k)},\qquad \frac{\mu (k,\tau )%
}{\lambda (k,\tau )+\mu (k,\tau )}=\frac{\nu _{\mu }(k)}{\nu _{\lambda
}(k)+\nu _{\mu }(k)}.
\end{equation*}%
In this case it is convenient to introduce the probabilities of jumping to
the right%
\begin{equation}
p_{\lambda }(k)=\frac{\nu _{\lambda }(k)}{\nu _{\lambda }(k)+\nu _{\mu }(k)}
\label{pr1}
\end{equation}%
and to the left%
\begin{equation}
p_{\mu }(k)=\frac{\nu _{\mu }(k)}{\nu _{\lambda }(k)+\nu _{\mu }(k)}.
\label{pr2}
\end{equation}%
Note that these jump probabilities are completely determined by the
anomalous exponents $\nu _{\lambda }(k)$ and $\nu _{\mu }(k).$ In the
standard CTRW theory, these jump probabilities are given independently \cite%
{Klages,MFH}.

Let us consider the non-local model for which the jump probabilities $%
p_{\lambda }(k)$ and $p_{\mu }(k)$ depend on the chemotactic substance $S(k)$
as follows
\begin{equation}
p_{\lambda }(k)=Ae^{-\beta \left( S(k+1)-S(k)\right) },\qquad p_{\mu
}(k)=Ae^{-\beta \left( S(k-1)-S(k)\right) },  \label{nonlocal}
\end{equation}%
where the parameter $A$ is determined from $p_{\lambda }(k)+$ $p_{\mu
}(k)=1. $ These jump probabilities describe the bias of cells with respect
to the spatial gradient $S(k+1)-S(k)$ \cite{Baker1,OS1}. One can obtain \cite%
{HL}
\begin{equation}
p_{\lambda }(k)-p_{\mu }(k)=\frac{e^{-\beta S(k+1)}-e^{-\beta S(k-1)}}{%
e^{-\beta S(k+1)}+e^{-\beta S(k-1)}}.  \label{diff}
\end{equation}

The transition PDF's $\psi _{\lambda }(k,\tau )=\lambda (k,\tau )\Psi
(k,\tau )$ and $\psi _{\mu }(k,\tau )=\mu (k,\tau )\Psi (k,\tau )$ can be
rewritten in terms of $\psi (k,\tau ),p_{\lambda }(k)$ and $p_{\mu }(k)$ as
\begin{equation}
\psi _{\lambda }(k,\tau )=p_{\lambda }(k)\psi (k,\tau ),\qquad \psi _{\mu
}(k,\tau )=p_{\mu }(k)\psi (k,\tau ).  \label{new3}
\end{equation}%
The asymptotic approximation for the Laplace transform of the waiting time
density $\psi (k,\tau )$ of the Pareto form (\ref{Pareto})\ can be found
from the Tauberian theorem \cite{Feller}
\begin{equation*}
\hat{\psi}\left( k,s\right) \simeq 1-g(k)s^{\nu (k)},\qquad s\rightarrow 0
\end{equation*}%
with%
\begin{equation}
g(k)=\Gamma (1-\nu (k))\left( \tau _{0}(k)\right) ^{\nu (k)}.
\end{equation}%
We obtain from (\ref{new5})\ and (\ref{new3}) the Laplace transforms of the
memory kernels%
\begin{equation}
\hat{K}_{\lambda }\left( k,s\right) \simeq \frac{p_{\lambda }(k)s^{1-\nu (k)}%
}{g(k)},\qquad \hat{K}_{\mu }\left( k,s\right) \simeq \frac{p_{\mu
}(k)s^{1-\nu (k)}}{g(k)},\qquad s\rightarrow 0.
\end{equation}%
Therefore, the integral escape rates to the right $i_{\lambda }(k,t)$ and to
the left $i_{\mu }(k,t)$ in the subdiffusive case are
\begin{eqnarray}
i_{\lambda }(k,t) &=&a(k)e^{-\theta (k)t}\mathcal{D}_{t}^{1-\nu (k)}\left[
p(k,t)e^{\theta (k)t}\right] ,  \notag \\
i_{\mu }(k,t) &=&b(k)e^{-\theta (k)t}\mathcal{D}_{t}^{1-\nu (k)}\left[
p(k,t)e^{\theta (k)t}\right] .
\end{eqnarray}%
Here $D_{t}^{1-\nu (k)}$ is the Riemann-Liouville fractional derivative with
varying order defined by (\ref{RL}). The anomalous rate functions $a(k)$\
and $b(k)$ are
\begin{eqnarray}
a(k) &=&\frac{p_{\lambda }(k)}{g(k)}=\frac{\nu _{\lambda }(k)}{\nu (k)\Gamma
(1-\nu (k))\left( \tau _{0}(k)\right) ^{\nu (k)}},\qquad  \notag \\
b(k) &=&\frac{p_{\mu }(k)}{g(k)}=\frac{\nu _{\mu }(k)}{\nu (k)\Gamma (1-\nu
(k))\left( \tau _{0}(k)\right) ^{\nu (k)}}  \label{ab}
\end{eqnarray}%
with the anomalous exponent $\nu (k)$ defined in (\ref{exp}).The master
equation (\ref{master99}) takes the form of non-homogeneous fractional
equation
\begin{eqnarray}
\frac{\partial p(k,t)}{\partial t} &=&a(k-1)e^{-\theta (k-1)t}\mathcal{D}%
_{t}^{1-\nu (k-1)}\left[ p(k-1,t)e^{\theta (k-1)t}\right]  \notag \\
&&+b(k+1)e^{-\theta (k+1)t}p(k+1)\mathcal{D}_{t}^{1-\nu (k+1)}\left[
p(k+1,t)e^{\theta (k+1)t}\right]  \notag \\
&&-\left( a(k)+b(k)\right) e^{-\theta (k)t}\mathcal{D}_{t}^{1-\nu (k)}\left[
p(k,t)e^{\theta (k)t}\right] -\theta (k)p(k,t)  \label{fractional}
\end{eqnarray}%
for $k>1.$

For $k=1$ with $\theta (k)=\chi =0$, we obtain%
\begin{equation}
\frac{\partial p(1,t)}{\partial t}=b(2)\mathcal{D}_{t}^{1-\nu (2)}p(2,t)-a(1)%
\mathcal{D}_{t}^{1-\nu (1)}p(1,t).  \label{f1}
\end{equation}%
The fractional probability flux $I_{\nu }(k,t)$ from the point $k$ to $k+1$
is \
\begin{equation}
I_{\nu }=a(k)e^{-\theta (k)t}\mathcal{D}_{t}^{1-\nu (k)}\left[
p(k,t)e^{\theta (k)t}\right] -b(k+1)e^{-\theta (k+1)t}p(k+1)\mathcal{D}%
_{t}^{1-\nu (k+1)}\left[ p(k+1,t)e^{\theta (k+1)t}\right] .  \label{If}
\end{equation}%
The equation (\ref{fractional}) can be rewritten in terms of the probability
flux $I_{\nu }(k,t)$ as
\begin{equation}
\frac{\partial p(k,t)}{\partial t}=-I_{\nu }(k,t)+I_{\nu }(k-1,t)-\theta
(k)p(k,t).
\end{equation}

\subsection{Fractional Fokker-Planck equation for cells density and
chemotaxis}

In this subsection we consider the continuous case ($k\rightarrow x$) and
find the drift $l(a(x)-b(x))$ together with diffusion coefficient in the
fractional Fokker-Planck equation (\ref{FFP}). It follows from (\ref{ab})
that the drift is proportional to the difference in the anomalous exponents $%
\nu _{\lambda }(x)$ and $\nu _{\mu }(x)$, since%
\begin{equation}
a(x)-b(x)=\frac{p_{\lambda }(x)-p_{\mu }(x)}{g(x)}=\frac{\nu _{\lambda
}(x)-\nu _{\mu }(x)}{\nu (x)\Gamma (1-\nu (x))\left( \tau _{0}(x)\right)
^{\nu (x)}}.  \label{dif1}
\end{equation}%
The difference $\nu _{\lambda }(x)-\nu _{\mu }(x)$ can be approximated in
the different ways. In the case of chemotaxis this difference is
proportional to the gradient of the local concentration of the chemotactic
substance $S(x).$ Using (\ref{diff}), we obtain
\begin{equation*}
p_{\lambda }(x)-p_{\mu }(x)=\frac{e^{-\beta S(x+l)}-e^{-\beta S(x-l)}}{%
e^{-\beta S(x+l)}+e^{-\beta S(x-l)}}.
\end{equation*}%
In the limit $l\rightarrow 0$, we have the standard chemotaxis model
\begin{equation}
a(x)-b(x)=\frac{p_{\lambda }(x)-p_{\mu }(x)}{g(x)}=-\frac{\beta l}{g(x)}%
\frac{\partial S}{\partial x}+o\left( l\right) ,  \label{dif2}
\end{equation}%
where the case \ $\beta >0$ corresponds to the negative taxis: the drift is
in the direction of the decrease in the value of the chemotactic substance $%
S(x)$. The fractional Fokker-Planck equation (\ref{FFP}) takes the form%
\begin{equation}
\frac{\partial \rho (x,t)}{\partial t}=\frac{\partial }{\partial x}\left[
\frac{l^{2}\beta }{g(x)}\frac{\partial S}{\partial x}\mathcal{D}_{t}^{1-\nu
(x)}\rho (x,t)\right] +\frac{\partial ^{2}}{\partial x^{2}}\left[ \frac{l^{2}%
}{2g(x)}\mathcal{D}_{t}^{1-\nu (x)}\rho (x,t)\right] .  \label{newFFP}
\end{equation}%
\begin{figure}[tbp]
\includegraphics[scale=0.4]{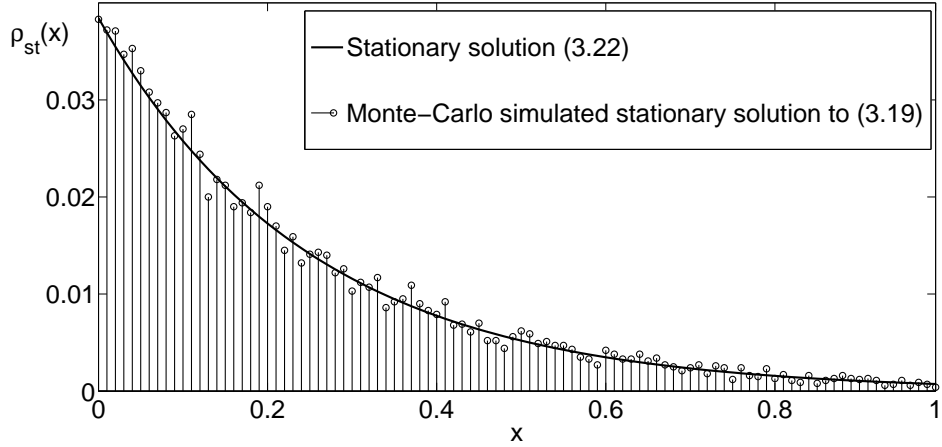}
\caption{{\ Monte-Carlo simulation of the stationary solution to equation (%
\protect\ref{newFFP}) and the analytical solution (\protect\ref{Bo}) for the
linear distribution of the chemotactic substance $S(x)=mx$ with $m=2$, $%
\protect\beta =10^{-2}.$}}
\label{fig1}
\end{figure}

As an example, let us consider the case when the anomalous exponent $\nu $
and time parameter $\tau _{0}$ are constants. Then the fractional
Fokker-Planck equation (\ref{newFFP}) can be rewritten as follows
\begin{equation}
\frac{\partial \rho (x,t)}{\partial t}=2\beta D_{\nu }\mathcal{D}_{t}^{1-\nu
}\frac{\partial }{\partial x}\left[ \frac{\partial S}{\partial x}\rho (x,t)%
\right] +D_{\nu }\mathcal{D}_{t}^{1-\nu }\frac{\partial ^{2}\rho (x,t)}{%
\partial x^{2}},  \label{Caputo2}
\end{equation}%
where $D_{\nu }$ is the fractional diffusion coefficient%
\begin{equation*}
D_{\nu }=\frac{l^{2}}{2\Gamma (1-\nu )\tau _{0}^{\nu }}.
\end{equation*}%
In the case of the reflective boundary conditions at $x=0$, the fractional
equation (\ref{Caputo2}) admits the stationary solution $\rho _{st}(x)$ in
the semi-infinite domain $[0,\infty ).$ It obeys the equation
\begin{equation}
2\beta \frac{\partial }{\partial x}\left[ \frac{\partial S(x)}{\partial x}%
\rho _{st}(x)\right] +\frac{\partial ^{2}\rho _{st}(x)}{\partial x^{2}}=0
\end{equation}%
and has the form of the Boltzmann distribution \cite{Met1,Met2}
\begin{equation}
\rho _{st}(x)=N^{-1}\exp \left[ -2\beta S(x)\right] ,  \label{Bo}
\end{equation}%
where $N=\int_{0}^{\infty }\exp \left[ -2\beta S(x)\right] dx.$ This
distribution describes the aggregation of cells due to nonuniform
distribution of the chemotactic substance $S(x).$ Fig. 1 illustrates the
stationary profile of cells density $\rho _{st}(x)=$ $2\beta m\exp \left[
-2\beta mx\right] $ for the linear distribution $S(x)=mx$ and $m=2$, $\beta
=10^{-2}$.

We use Monte-Carlo method to simulate a stationary solution to equation (\ref%
{newFFP}). We select the following parameters: $\nu =0.5$, $\tau _{0}=1$,
and $l=0.01$. The Fig. 1 shows the result of the $10^{6}$ simulated random
walk trajectories, with jump probabilities given by (\ref{nonlocal}), Pareto
waiting time distribution (\ref{Pareto}), and terminal time $T=10^{6}$.

In the next section we will be concerned with the asymptotic behavior of the
solution of the master equation $p(k,t)$ as $t\rightarrow \infty $ \ for $%
\theta (k)=0.$ In particular, we will show that the stationary distribution (%
\ref{Bo}) is not structurally stable with respect to the spatial variations
of the anomalous exponent.

\section{Structural instability and anomalous aggregation}

It has recently been shown that the subdiffusive fractional equations with
constant anomalous exponent $\nu $ in a bounded domain $\left[ 0,L\right] $\
are not structurally stable with respect to the non-homogeneous variations
of parameter $\nu $ \cite{Fed3}. It turns out that the spatial variations of
the anomalous exponent lead to a drastic change in asymptotic behavior of $%
p(k,t)$ for large $t.$ The purpose of this section is to find the conditions
of this structural instability in semi-infinite domain $1\leq k<\infty .$ We
consider the case when $\theta (k)=\chi =0$ for which the total probability
is conserved
\begin{equation}
\sum\limits_{k=1}^{\infty }p(k,t)=1  \label{conse}
\end{equation}%
and the fractional probability flux $I_{\nu }(k,t)$ from the point $k$ to $%
k+1$ is
\begin{equation}
I_{\nu }(k,t)=a(k)\mathcal{D}_{t}^{1-\nu (k)}p(k,t)-b(k+1)p(k+1)\mathcal{D}%
_{t}^{1-\nu (k+1)}p(k+1,t).  \label{If0}
\end{equation}%
For simplicity we assume that the initial conditions are $p_{0}(1)=1$ and $%
p_{0}(k)=0$ for $k\neq 1.$Taking the Laplace transform of (\ref{dis1}) and (%
\ref{conse}) we obtain
\begin{eqnarray}
s\hat{p}(k,s) &=&a(k-1)s^{1-\nu (k-1)}\hat{p}(k-1,s)+b(k+1)s^{1-\nu (k+1)}%
\hat{p}(k+1,s)  \notag \\
&&-\left( a(k)+b(k)\right) s^{1-\nu (k)}\hat{p}(k,s)  \label{L2}
\end{eqnarray}%
and%
\begin{equation}
\sum\limits_{k=1}^{\infty }s\hat{p}(k,s)=1.  \label{norm}
\end{equation}%
Since there is no flux of cells outside the left border, we have for $k=1$%
\begin{equation}
s\hat{p}(1,s)-1=b(2)s^{1-\nu (2)}\hat{p}(2,s)-a(1)s^{1-\nu (1)}\hat{p}(1,s).
\label{L1}
\end{equation}%
In the limit $s\rightarrow 0,$ one can obtain from (\ref{L1}) simple formula
expressing $\hat{p}(2,s)$ in terms of $\hat{p}(1,s)$%
\begin{equation*}
\hat{p}(2,s)\simeq \frac{a(1)s^{\nu (2)-\nu (1)}}{b(2)}\hat{p}(1,s),\qquad
s\rightarrow 0.
\end{equation*}%
In general, we find from (\ref{L2}) and (\ref{L1}) $\hat{p}(k,s)$ in terms
of $\hat{p}(k-1,s)$
\begin{equation}
\hat{p}(k,s)\simeq \frac{a(k-1)s^{\nu (k)-\nu (k-1)}}{b(k)}\hat{p}%
(k-1,s),\qquad k>1,\qquad s\rightarrow 0.  \label{re1}
\end{equation}%
This formula has very simple probabilistic meaning: the flux $I_{\nu
}(k-1,t)\rightarrow 0$ as $t\rightarrow \infty .$ If we take the Laplace
transform of $I_{\nu }(k-1,t)$ from (\ref{If0}), we obtain
\begin{equation}
\hat{I}_{\nu }(k-1,s)=a(k-1)s^{1-\nu (k-1)}\hat{p}(k-1,s)-b(k)p(k)s^{1-\nu
(k)}\hat{p}(k,s).
\end{equation}%
It follows from (\ref{re1}) that $\hat{I}_{\nu }(k,s)\simeq 0$ as $%
s\rightarrow 0.$

\subsection{ Stationary solution to fractional equation with constant
anomalous exponent}

Let us assume that the anomalous exponent $\nu (k)$ is independent of the
position $k$ that is $\nu =const.$ Let us find stationary solution to the
fractional master equation (\ref{dis1})%
\begin{equation}
p_{st}(k)=\lim_{t\rightarrow \infty }p(k,t)=\lim_{s\rightarrow 0}s\hat{p}%
(k,s).  \label{st}
\end{equation}%
It follows from (\ref{re1}) that
\begin{equation*}
\hat{p}(k,s)\simeq \frac{a(k-1)}{b(k)}\hat{p}(k-1,s),\qquad k>1,\qquad
s\rightarrow 0.
\end{equation*}%
or
\begin{equation}
\hat{p}(k,s)\simeq \prod\limits_{j=1}^{k-1}\frac{a(j)}{b(j+1)}\hat{p}%
(1,s),\qquad k>1,\qquad s\rightarrow 0.
\end{equation}%
Using the normalization condition (\ref{norm}) and (\ref{st}), we obtain the
stationary solution of the equation (\ref{dis1})%
\begin{equation}
p_{st}(k)=p_{st}(1)\prod\limits_{j=1}^{k-1}\frac{a(j)}{b(j+1)},\qquad k>1,
\label{an}
\end{equation}%
where
\begin{equation*}
p_{st}(1)=\left( 1+\sum\limits_{k=2}^{\infty }\prod\limits_{j=1}^{k-1}\frac{%
a(j)}{b(j+1)}\right) ^{-1}.
\end{equation*}%
If the sum
\begin{equation*}
\sum\limits_{k=2}^{\infty }\prod\limits_{j=1}^{k-1}\frac{a(j)}{b(j+1)}
\end{equation*}%
is divergent, the stationary solution does not exist. In particular, if we
assume that the anomalous rate functions $a$ and $b$ are equal, that is, $%
a(k)=b(k+1),$ then for a finite domain with $k=1,2,...,N,$ we obtain uniform
distribution $p_{st}(k)=1/N$ \ for every $k.$ The stationary solution (\ref%
{an}) is very similar to (\ref{ma}) corresponding to the Markov birth-death
model. However, this similarity is very deceptive, because (\ref{an}) is not
structurally stable with respect to the non-homogeneous variations of
parameter $\nu $. The aim of next subsection is to show this structural
instability.

\subsection{\ Anomalous aggregation.}

Now we consider non-homogeneous case for which the anomalous exponent
depends on $k.$ We assume that the point $k=M$ has the property that $\nu
(M)<\nu (k)$ for all $k\neq M$. Our purpose now is to find the conditions
under which%
\begin{equation}
\lim_{t\rightarrow \infty }p(M,t)=1,\qquad \lim_{t\rightarrow \infty
}p(k,t)=0,\qquad k\neq M.  \label{ag}
\end{equation}%
It means that the total probability concentrates just at one point $k=M.$
This phenomenon is called an anomalous aggregation \cite{Fed1}. This
asymptotic behavior of cells was observed in experiments on phagotrophic
protists when \textquotedblleft cells become immobile in attractive patches,
which will then eventually trap all cells\textquotedblright\ \cite{AA}. In
the Laplace space, (\ref{ag}) takes the form
\begin{equation}
\lim_{s\rightarrow 0}s\hat{p}(M,s)=1,\qquad \lim_{s\rightarrow 0}s\hat{p}%
(k,s)=0,\qquad k\neq M.  \notag
\end{equation}%
We can rewrite the normalization condition (\ref{norm}) as%
\begin{equation}
s\hat{p}(M,s)+\sum\limits_{k=1}^{M-1}s\hat{p}(M-k,s)+\sum\limits_{k=1}^{%
\infty }s\hat{p}(M+k,s)=1.  \label{cons}
\end{equation}%
By using (\ref{re1}), we express $\hat{p}(M+k,s)$ in terms of $\hat{p}(M,s)$%
\ as follows
\begin{equation}
\hat{p}(M+k,s)\simeq \hat{p}(M,s)\prod\limits_{j=1}^{k}\frac{a(M+j-1)}{b(M+j)%
}s^{\nu (M+k)-\nu (M)},\qquad k\geq 1,\qquad s\rightarrow 0.  \label{M1}
\end{equation}%
Now we write the formula for $\hat{p}(M-k,s)$ in terms of $\hat{p}(M,s)$
\begin{equation}
\hat{p}(M-k,s)\simeq \hat{p}(M,s)s^{\nu (M-k)-\nu (M)}\prod\limits_{j=1}^{k}%
\frac{b(M-j+1)}{a(M-j)},\qquad k=1,...,M-1,\qquad s\rightarrow 0.  \label{M2}
\end{equation}%
Now we substitute (\ref{M1}) and (\ref{M2}) into (\ref{cons}) and use $s\hat{%
p}(M,s)$ as a common factor
\begin{equation}
s\hat{p}(M,s)\left( 1+\sum\limits_{k=1}^{M-1}s^{\nu (M-k)-\nu
(M)}\prod\limits_{j=1}^{k}\frac{b(M-j+1)}{a(M-j)}+\sum\limits_{k=1}^{\infty
}s^{\nu (M+k)-\nu (M)}\prod\limits_{j=1}^{k}\frac{a(M+j-1)}{b(M+j)}\right)
\simeq 1.  \notag
\end{equation}%
Since $\nu (M)<\nu (k)$ for any $k\neq M$, we have $s^{\nu (M+k)-\nu
(M)}\rightarrow 0$ and $s^{\nu (M-k)-\nu (M)}\rightarrow 0$ as $s\rightarrow
0.$ We conclude that if%
\begin{equation}
\sum\limits_{k=1}^{\infty }s^{\nu (M+k)-\nu (M)}\prod\limits_{j=1}^{k}\frac{%
a(M+j-1)}{b(M+j)}\rightarrow 0  \notag
\end{equation}%
as $s\rightarrow 0,$ then $s\hat{p}(M,s)\rightarrow 1,$ while $s\hat{p}%
(k,s)\rightarrow 0$.\ It means that in the limit $t\rightarrow \infty ,$ we
obtain (\ref{ag}). If instead of probability $p(k,t)$ we consider the mean
density of cells $\rho \left( x,t\right) ,$ the formula (\ref{ag}) can be
rewritten as $\rho \left( x,t\right) \rightarrow \delta (x-x_{\min })$ as $%
t\rightarrow \infty ,$\ where $x_{\min }$ is the point on the interval $%
[0,\infty )$ at which $\nu (x)$ takes its minimum value. Note that this
result was obtained for a symmetrical random walk in the context of
chemotaxis and anomalous aggregation \cite{Fed1}.

\section{Conclusions.}

We have studied a non-homogeneous in space and non-local in time random walk
model describing anomalous subdiffusive transport of cells. Using a Markov
model with structured probability density function, we have derived
non-local in time and fractional master equations for the probability of
cell position. The advantage of our probabilistic approach is that it allows
us to take into account the death process within the general non-Markovian
random walk. The main feature of our fractional model is that the transition
probabilities for jumping on the left and right depend inversely on the
residence time variable. This dependence induces power-law residence time
distribution and ultimately the anomalous subdiffusion of cells. It has
recently been shown that the subdiffusive fractional equations with constant
anomalous exponent\ are not structurally stable in a bounded domain with
respect to the non-homogeneous variations. In this paper we have extended
and complemented our previous results for infinite domain and found exact
conditions under which the structural instability takes place. Our model can
be generalized in many ways, e.g., by modelling the residence time by
internal chemical reactions via a stochastic or ordinary differential
equations instead of simple equation for the residence time $d\tau /dt=1$.
It would be interesting to take into account the density-dependent dispersal
\cite{Campos2} including non-linear exclusion process with cell-to-cell
adhesion \cite{Baker, Khain}.

\section*{Acknowledgements}

The authors gratefully acknowledge the support of the Federal Programme
N 14.A18.21.0867. SF acknowledges the warm hospitality of Department
of Mathematical Physics, Ural Federal University. SF also acknowledges the
support of the EPSRC Grant EP/J019526/1. The authors wish to thank S.
Falconer for interesting discussions.


\end{document}